\documentclass{ceab}   

\usepackage{epsfig}     
\usepackage{graphicx}   
\usepackage{url}
\usepackage{natbib}     
\usepackage[T1]{fontenc} 
    
\usepackage{lmodern}
\usepackage{pgf}
\usepackage{tikz}
\usetikzlibrary{decorations.pathreplacing}
\usepackage{xcolor}
\usepackage{xspace}
\usepackage{hyperref}

\newcommand{\corsika}{\textsc{Corsika}\xspace}
\newcommand{\pythia}{\textsc{Pythia}\xspace}

\usepackage[english]{babel}
\usepackage{rotating}
\usepackage{float}

\def\runningtitle{Gaudu et al.: Status of the \corsika 8 particle-shower simulation code} 
\def\articlenumber{7} 
\def\ppages{1--10} 

\begin{document}

\title{Status of the CORSIKA 8 particle-shower simulation code}

\author{C. Gaudu$^{1}$ for the CORSIKA 8 Collaboration\footnote{\scriptsize FAL: \href{https://gitlab.iap.kit.edu/AirShowerPhysics/corsika/-/wikis/Current-CORSIKA-8-author-list}{gitlab.iap.kit.edu/AirShowerPhysics/corsika/-/wikis/Current-CORSIKA-8-author-list}}
\vspace{2mm}\\
\it $^1$Bergische Universität Wuppertal,\\ 
\it Gaußstraße 20, 42119 Wuppertal, Germany,\\
}

\maketitle

\begin{abstract}
For over two decades, \corsika 7 and its previous versions have been the leading Monte Carlo code for simulating extensive air showers. However, its monolithic Fortran-based software design and hand-optimized code has created challenges for maintenance, adaptation to new computing paradigms, and extensions for more complex simulations. Addressing these limitations, the \corsika 8 project represents a comprehensive rewrite of \corsika 7, seeing its core functionality re-envisioned in a modern and modular C++ framework.

\corsika 8 has now reached a ``physics-complete'' state, offering a robust platform that encourages expert development for specialized applications. It supports high-energy hadronic interactions using models such as \textsc{Sibyll} 2.3d, \textsc{QGSJet}-II.04, EPOS-LHC, and \pythia 8.3, alongside the treatment of electromagnetic cascades with \textsc{PROPOSAL} 7.6.2. Key highlights are the support for multiple interaction media, including cross-media particle showers, and an enhanced calculation of radio emissions from particle showers.

This contribution provides an overview of the current functionalities, showcases validation results of its simulations, and discusses future development plans.
\end{abstract}


\section{Introduction}

The astroparticle physics community significantly depends on Monte Carlo simulations to analyze particle showers occurring in air and various other media. For over two decades, the \corsika code (\cite{Heck:1998vt}) has served as the standard for air shower simulations, initially developed for the KASCADE experiment. Currently maintained by KIT, \corsika 7.75 faces challenges due to its monolithic Fortran architecture and the retirement of key developers, complicating ongoing maintenance efforts. Additionally, forthcoming experiments necessitate a level of flexibility that the hand-optimized \corsika~7 (C7) code cannot accommodate in the long run. 

Consequently, in 2018, the development of \corsika 8 (C8) was initiated, which entails a complete rewrite of the core functionalities of C7 within a modular C++-based framework. The design philosophy behind C8 was previously discussed in~\cite{CORSIKA:2023jyz}, alongside with the general structure of the code. By the time of ICRC 2023, the \corsika 8 code was deemed ``physics-complete'' and since then, the development was focused primarly on enhancing the user experience. 

This work reviews the most recent results in terms of electromagnetic and hadronic cascades, radio emission calculations and discuss the usage of \corsika 8 by the physics community.

\section{Electromagnetic cascades}

Electromagnetic cascades in \corsika 8 are treated using the PROPOSAL code (\cite{Alameddine_2020}) at current version 7.6.2. In~\cite{Sandrock:2023nfn}, an overview of the simulation of electromagnetic cascades within \corsika 8 is presented, with an emphasis on electron-induced showers. It includes a comparative analysis of the implemented physics and simulation outcomes with those produced by EGS4, as is the case in \corsika 7~\footnote{It is important to note that although \corsika 7 serves as a benchmark for comparison, certain physical processes have been enhanced and differ in \corsika 8; e.g. C8 includes the process of triplet production, which is not accounted for by EGS4 in C7. A 1:1 agreement with C7 is neither expected nor intended.}. 

\begin{figure}[h!] 
    \centering
    \includegraphics[width=0.52\linewidth]{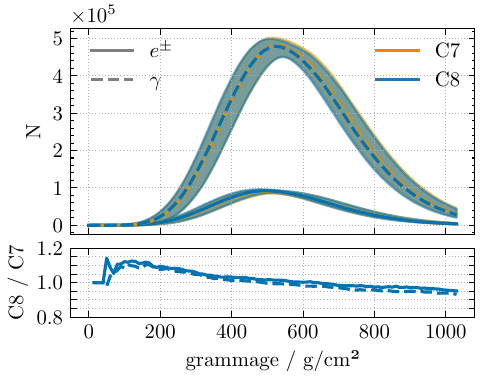}
    \caption{Longitudinal profile of electromagnetic particles for 100 TeV photon-induced showers, observed at a height of 5.8 km above sea level, in \corsika 7 and \corsika 8.}
    \label{fig:longitudinal_em_profile}
\end{figure}

In this section, we display results of the electromagnetic component of \textit{photon}-induced air showers. The longitudinal profiles in Fig.~\ref{fig:longitudinal_em_profile} shows a consistency between \corsika 7 and \corsika 8 on the few-percent level, except for very low particle numbers at the earliest stage of the shower development.

\begin{figure}[h!] 
    \hspace{-0.4cm}
    \includegraphics[width=0.52\linewidth]{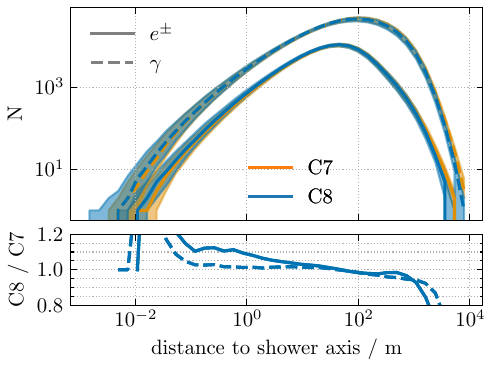}
    \includegraphics[width=0.52\linewidth]{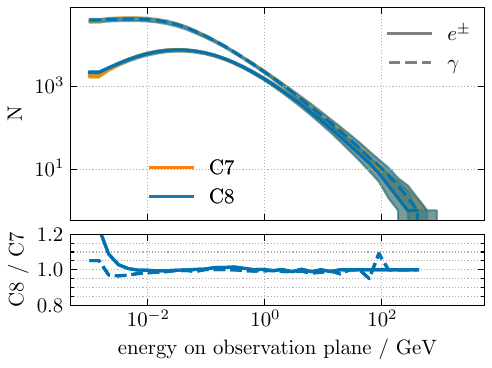}
    \vspace{-0.4cm}\caption{Lateral profile (left) and energy distribution on the observation plane (right) of electromagnetic particles for 100 TeV photon-induced showers, observed at a height of 5.8 km above sea level, in \corsika 7 and \corsika 8.}
    \label{fig:ldf_espec_em_profile}
\end{figure}

The lateral distributions and energy spectra are shown in Fig.~\ref{fig:ldf_espec_em_profile}. The lateral distributions find themselves in agreement at the 10\% level for e$^\pm$, while the codes agree below the 5\% level for photons, except for very close distances to the shower core, at the cm scale, where both C8 distributions diverge from the C7 description. The energy spectra for C8 and C7 are in agreement typically on a 5\% level, with some more pronounced differences at very low energies. 

At extremely high energies and also at large densities, the Landau-Pomeranchuk-Migdal effect changes the behavior of electromagnetic showers, slowing down the development of the cascade and shifting the shower maximum to larger depth, as discussed in~\cite{Sandrock:2023nfn}.

\section{Hadronic cascades}

\corsika 8 employs a modular framework that integrates state-of-the-art physics models to simulate hadronic processes with precision and adaptability. It supports a versatile suite of high-energy interaction models, including EPOS-LHC (\cite{Pierog:2013ria}), \textsc{Sibyll} 2.3d (\cite{Riehn:2019jet}), \textsc{QGSJet}-II.04 (\cite{Ostapchenko:2010vb}), and an experimental integration of \pythia 8.3 (\cite{Bierlich:2022pfr}), catering to diverse physics scenarios. Most recent efforts on the integration of the \pythia 8.3 hadronic interaction model are discussed in~\cite{Gaudu:2024mkp} and~\cite{Gaudu:2024rsq}, while in parallel, ongoing work to interface the recently updated models, EPOS LHC-R (\cite{Pierog:2023ahq}) and \textsc{QGSJet}-III (\cite{Ostapchenko:2024onp}) is taking place. 

The model used in \corsika 8 for low-energy interactions is FLUKA (\cite{Ballarini:2024uxz}). The transition threshold between the high- and low-energy models is fixed at $\sim 80$~GeV as default, but can be modified. Decay processes are supported through \textsc{Sibyll} 2.3d and \pythia~8.

\begin{figure}[h!]
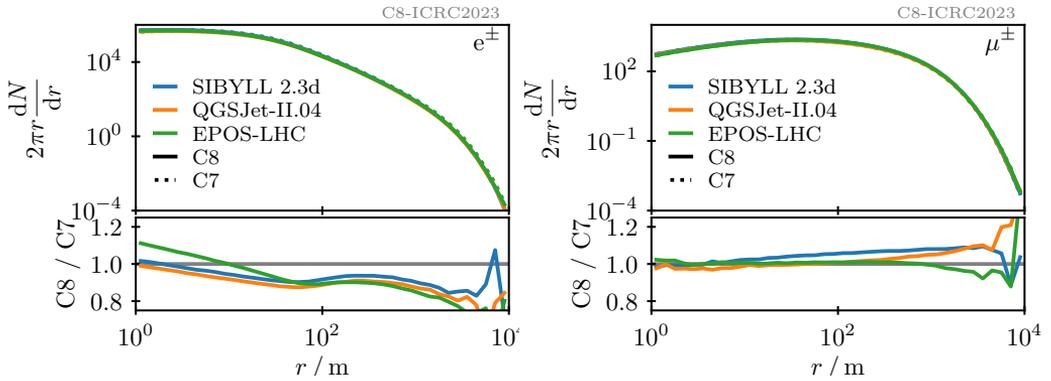
 
    \hspace{-0.8cm}
    \resizebox{0.55\linewidth}{!}{\input{fig/ldf_e+-.pgf}}
    \hspace{-0.3cm}
    \resizebox{0.55\linewidth}{!}{\input{fig/ldf_mu+-.pgf}} 
    \begin{tikzpicture}[remember picture, overlay]
        \node (label1) at (5,5.25) {};  
        \node (label2) at (11.75,5.25) {};
        
        \node[font=\rmfamily] at (label1) {\tiny \textcolor{gray}{C8-ICRC2023}};
        \node[font=\rmfamily] at (label2) {\tiny \textcolor{gray}{C8-ICRC2023}};
    \end{tikzpicture}
    \vspace{-0.6cm} \caption{Average lateral distributions of electrons/positrons (left) and muons (right) at ground for 300 vertical proton-induced 10$^{17}$ eV air showers, using several high-energy hadronic interaction models and FLUKA as low-energy interaction model, for \corsika 8 and \corsika 7. From~\cite{CORSIKA:2023jyz}.}
    \label{fig:ldf}
\end{figure}

In Fig.~\ref{fig:ldf}, comparisons of the lateral distributions of electrons/positrons and muons at the ground are displayed for proton-induced showers with the interaction models EPOS-LHC, \textsc{Sibyll} 2.3d, \textsc{QGSJet}-II.04. Reaching agreement at this level with completely independent codes is no small accomplishment, though further investigation is needed to understand the differences in the lateral distribution of electrons and positrons in hadron-induced air showers between C7 and C8. The muon content at ground agrees at the 5–10\% level near the shower axis, while discrepancies at larger distances remain to be explored. 

\section{Radio-emission calculations}

The radio-emission calculations have significantly influenced the development of \corsika 8. This is especially true given the radio-detection community's requirement for a flexible framework that can accommodate complex simulation scenarios. For instance, in-ice radio detection experiments necessitate consideration of the radio emission from air showers as they propagate from air into ice, a background that is presently simulated by piecing together several simulation codes (\cite{DeKockere:2024qmc}). The implementation of the radio process in \corsika 8 is discussed in~\cite{CORSIKA:2023soz} and most recently by~\cite{Alameddine:2024cyd}.

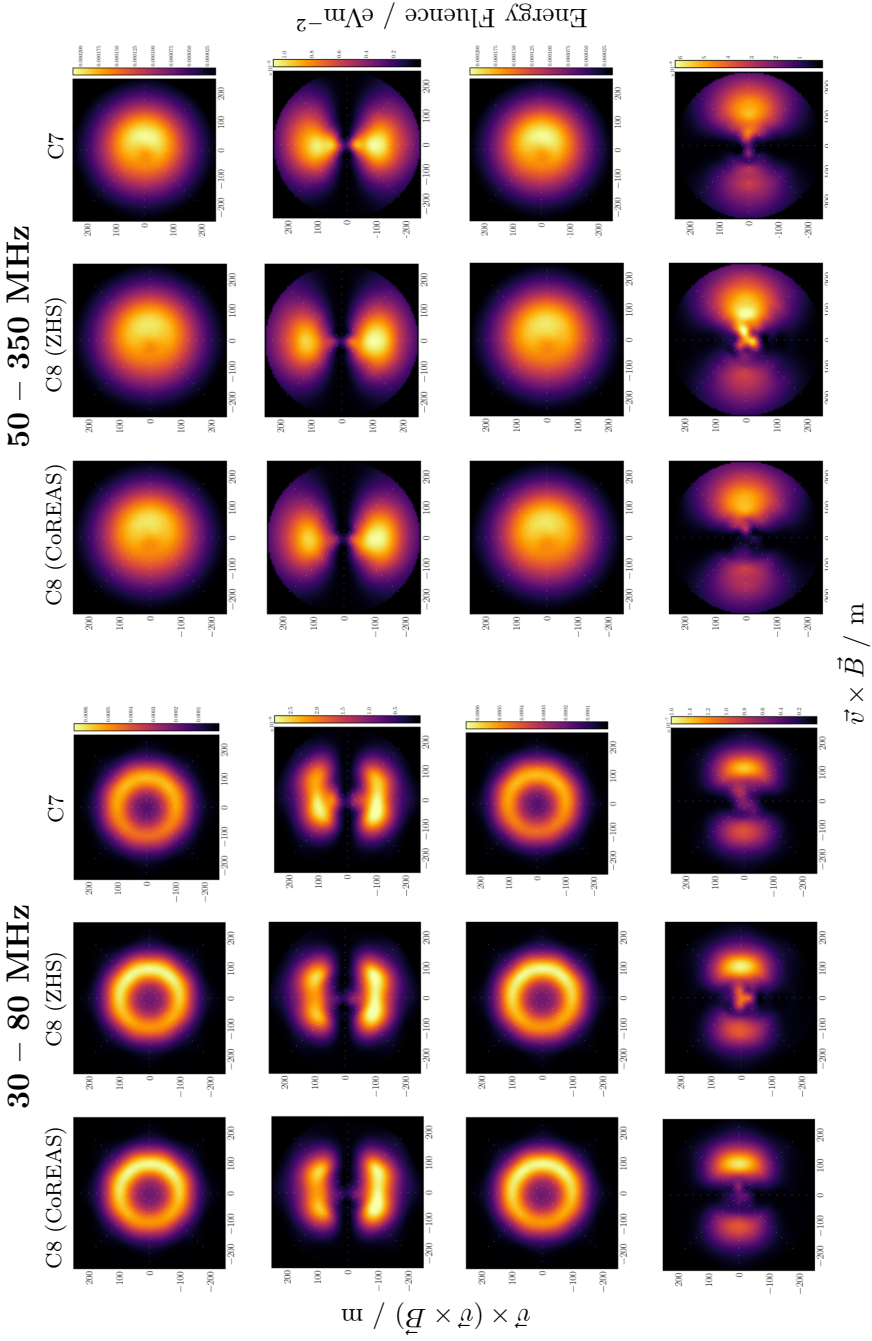
\begin{sidewaysfigure} \hspace{0.5cm}
    \includegraphics[width=0.45\linewidth, trim={10cm 0.7cm 10cm 0cm}, clip]{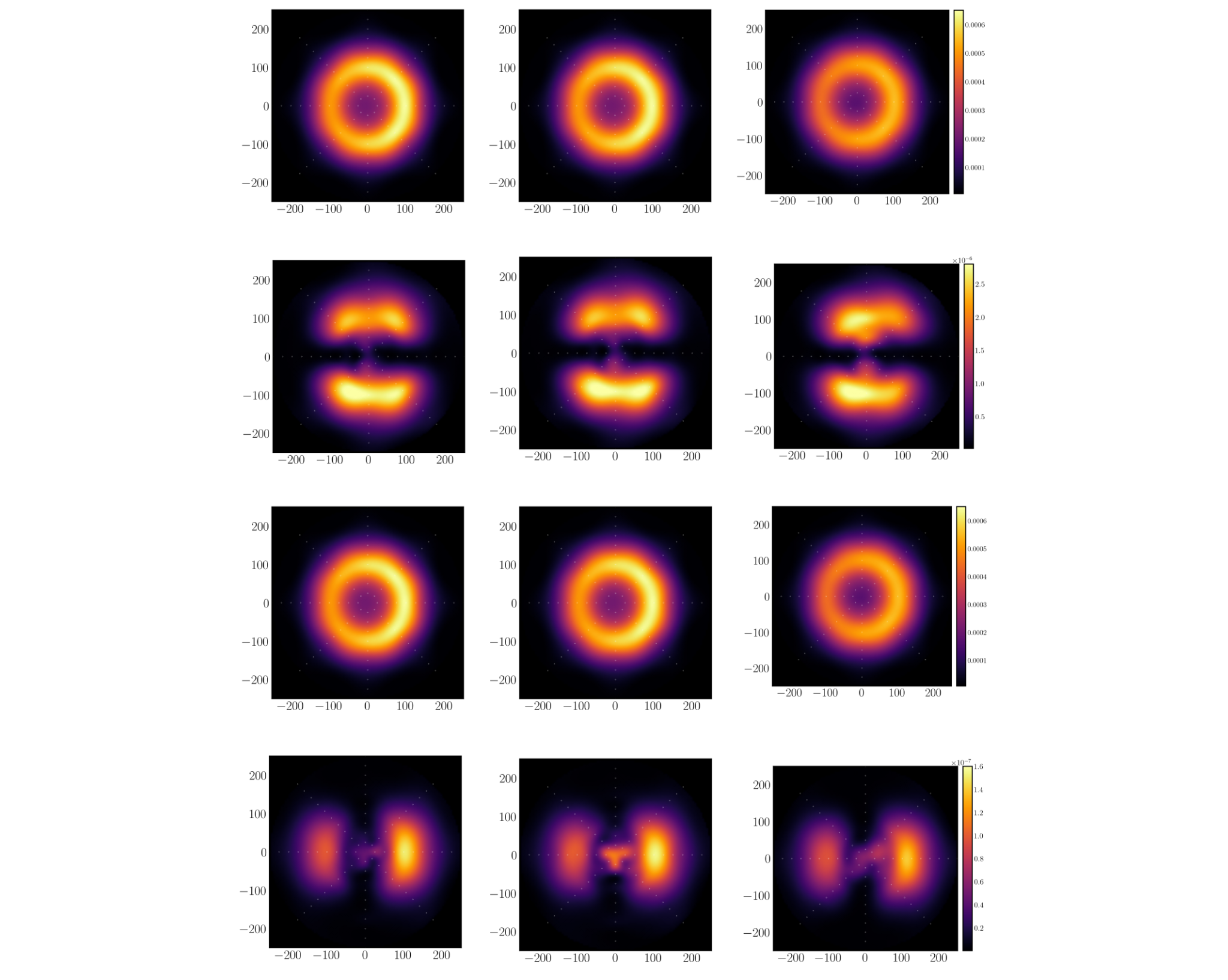} \hspace{0.5cm}
    \includegraphics[width=0.45\linewidth, trim={10cm 0.7cm 10cm 0cm}, clip]{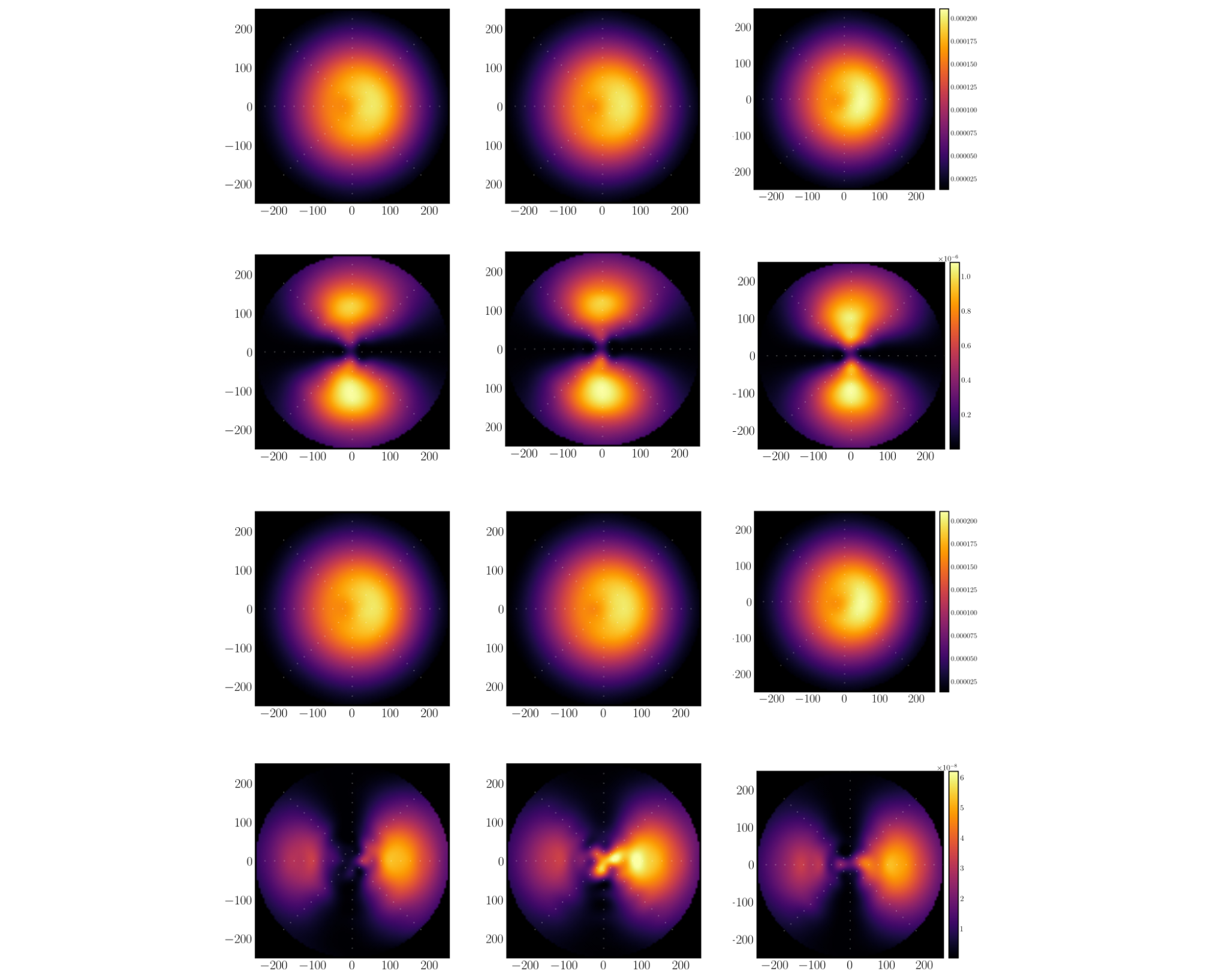} 
    \begin{tikzpicture}[remember picture, overlay]
        \node (label1) at (-15.5, 10.1) {}; 
        \node (label2) at (-12.9, 10.1) {};
        \node (label3) at (-10.4, 10.1) {};
        \node (label1b) at (-6.8, 10.1) {};
        \node (label2b) at (-4.3, 10.1) {};
        \node (label3b) at (-1.7, 10.1) {};
        \node (label4) at (-13, 10.6) {};
        \node (label5) at (-4.2, 10.6) {};
        
        \node[font=\rmfamily] at (label1) {\scriptsize C8 (CoREAS)};
        \node[font=\rmfamily] at (label2) {\scriptsize C8 (ZHS)};
        \node[font=\rmfamily] at (label3) {\scriptsize C7};
        \node[font=\rmfamily] at (label1b) {\scriptsize C8 (CoREAS)};
        \node[font=\rmfamily] at (label2b) {\scriptsize C8 (ZHS)};
        \node[font=\rmfamily] at (label3b) {\scriptsize C7};

        \node[font=\rmfamily] at (label4) {\large \textbf{30 -- 80 MHz}};
        \node[font=\rmfamily] at (label5) {\large \textbf{50 -- 350 MHz}};
        
        \node[anchor=north, font=\rmfamily] at (-8.5, 0) {\small $\vec{v} \times \vec{B}$ / m};
        \node[anchor=east, rotate=90, font=\rmfamily] at (-17.1, 6.5) {\small $\vec{v} \times (\vec{v} \times \vec{B})$ / m};
        \node[anchor=west, rotate=90, font=\rmfamily] at (0, 3) {\small Energy Fluence / eVm$^{-2}$};
    \end{tikzpicture}
    \vspace{0.3cm}
    \caption{Fluence maps for 30-80 MHz (left) and 50-350 MHz (right) radio emission from a vertical 10$^{17}$ eV iron-induced air shower simulated with both the endpoint (CoREAS) and ZHS formalisms in \corsika 8 compared with CoREAS from \corsika~7, for smaller step size (MaxRad $\sim 0.001$ rad). From top to bottom, the rows show: total energy fluence, fluence in the $\vec{v} \times (\vec{v} \times \vec{B})$ (north-south), $\vec{v} \times \vec{B}$ (east-west) and $\vec{v}$ (vertical) polarizations. The $x/y$-axes show core distances $\vec{v} \times \vec{B}$ and $\vec{v} \times (\vec{v} \times \vec{B})$ directions. The color scales are identical within a given row.}
    \label{fig:radio}
\end{sidewaysfigure}

The best method to examine the agreement for all observer positions between \corsika 7 (C7), \corsika 8 (C8) and \textsc{ZHAires} (\cite{Zas:1991jv}) is with the energy fluence, which is the energy deposited by radio waves per unit area in a given frequency band, predicted at the ground. Displayed in Fig.~\ref{fig:radio} is the 2D fluence map for different signal polarizations of the electric field in the 30 MHz to 80 MHz, and 50 MHz to 350 MHz frequency bands in the shower plane, for small step sizes (MaxRad $\sim$ 0.001 rad), for C8 with ``CoREAS'' formalism, C8 with ``ZHS'' formalism, and C7 with CoREAS extension.

The emission footprints from \corsika 8 with both formalism, and \corsika 7, demonstrate an agreement in terms of shape and symmetry of the energy fluence for all of the electric field components. In~\cite{Alameddine:2024cyd}, it is shown that as the particle track length in the simulation is decreased, the ``CoREAS'' and ``ZHS'' formalisms agree to within 2\% in the 30–80 MHz band and to within 1\% in the 50–350 MHz band. For small particle tracks, the agreement in terms of radiation energy, which is the area integral over the energy fluence, is on the level of better than 10\% for the 30-80 MHz band.

An interesting feature of the vertical polarization is the presence of a \emph{blip} near the shower axis predicted by the ``ZHS'' formalism in both C8 and \textsc{ZHAireS}\footnote{This \emph{blip} can be spotted for C8(ZHS) in the bottom row of Fig.\ \ref{fig:radio}, and in~\cite{Alameddine:2024cyd} for C8(ZHS) and \textsc{ZHAireS}.}, a feature that is not observed in the calculations employing the ``CoREAS'' formalism. From a practical point of view, however, the polarization along the shower axis holds little relevance, as the fluence in that region is negligible. 

\section{New functionalities}

Certain functionalities introduced in \corsika 8 cannot be independently achieved using its predecessor, \corsika 7. It is now feasible to trace the ancestry of particles beyond the immediate mother and grandmother generations, enabled by the newly introduced air shower genealogy feature, as outlined in~\cite{Reininghaus:2021zge}. Furthermore, the thinning algorithm for electromagnetic cascades has been enhanced by incorporating elements of the Hillas algorithm, previously utilized in \corsika 7, alongside a statistical thinning approach. More details about this updated thinning algorithm are discussed in~\cite{CORSIKA:2023jyz}.  

\begin{figure}[h!]
    \centering
    \includegraphics[width=0.55\linewidth, trim={0.1cm 0.35cm 0.1cm 0.33cm}, clip]{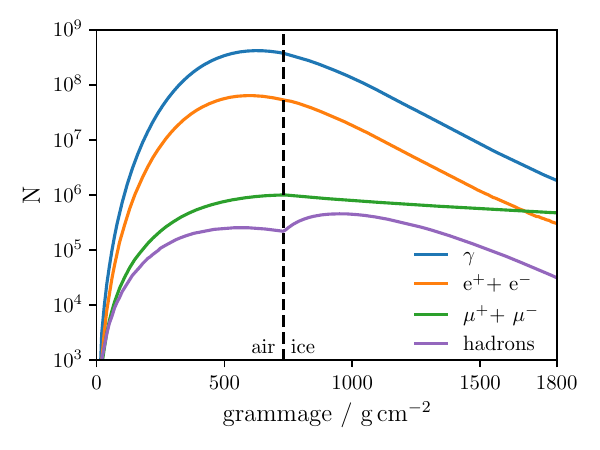}
    \vspace*{-0.1cm}\caption{Longitudinal profile of a vertical 100 PeV proton \emph{cross-media} shower, starting at the top of the atmosphere, and intersecting an ice layer located at an altitude of 2.4 km above sea level. Data from~\cite{CORSIKA:2023thf}.}
    \label{fig:new_feature}
\end{figure}

A most-awaited implementation for the in-ice radio-detection community is the ability to simulate showers propagating through multiple media \emph{within a single framework}, now made possible by the cross-media shower feature in \corsika 8. Combining this novel feature with radio emission calculations will allow the study of complex geometries relevant for radio neutrino detectors. Recent advancements in simulating in-ice radio emissions from air showers using \corsika 8 are presented in~\cite{Coleman:2024cln}. This study shows, within the single framework of C8, that air-shower cores generate coherent in-ice radio emissions similar to those generated by neutrino-induced in-ice showers of equivalent energy. This represents an important step forward for the in-ice radio community, expanding the range of available tools.

\section*{Conclusion}

There were many improvements over the last two years, and we now stand at a stage where the \corsika 8 code is considered ``physics-complete''. This means the implementation of the physical processes taking place during the development of particle showers is done. 

Recent enhancements include the implementation of the low-energy interaction model FLUKA, ongoing efforts to integrate the high-energy interaction model \pythia 8.3, consideration of photo-hadronic interactions and the Landau-Pomeranchuk-Migdal (LPM) effect, refinement of the electromagnetic cascade thinning algorithm, a fully integrated Cherenkov light calculation using GPU parallelization (\cite{CORSIKA:2023sfd}), and a realistic simulation of radio emission in air accompanied by a proof-of-concept CPU parallelization (\cite{CORSIKA:2023nqx}). Extensive validation was done against \corsika 7 and other codes; what remains is understanding differences at the 5-10\% level, polishing user-level and technical aspects, such as refining simulation steering, enhancing performance, and finalizing the documentation. An expert release of \corsika 8 is expected early 2025.

A future implementation of interaction models \textsc{QGSJet}-III and EPOS LHC-R to \corsika 8 is shortly expected. Moreover, additional features are anticipated as user requests are communicated to the development team. Users interested in testing \corsika 8 for specific use cases are encouraged to contact the development team through the channels provided on the Corsika 8 GitLab page\footnote{\href{https://gitlab.iap.kit.edu/AirShowerPhysics/corsika}{gitlab.iap.kit.edu/AirShowerPhysics/corsika}}.

\section*{Acknowledgements} 
C. Gaudu acknowledges funding from the German Research Foundation (DFG, Deutsche Forschungsgemeinschaft) via the Collaborative Research Center SFB1491: Cosmic Interacting Matters – from Source to Signal (F4) -- project no. 445052434.

\newpage

\bibliographystyle{apalike}
\bibliography{corsika_8}

\begin{thebibliography}{}

\bibitem[Alameddine et~al., 2023a]{CORSIKA:2023soz}
Alameddine, J.-M. et~al. (2023a).
\newblock {Simulating radio emission from air showers with CORSIKA 8}.
\newblock {\em PoS}, ICRC2023:425.

\bibitem[Alameddine et~al., 2023b]{CORSIKA:2023thf}
Alameddine, J.-M. et~al. (2023b).
\newblock {Simulations of cross media showers with CORSIKA 8}.
\newblock {\em PoS}, ICRC2023:442.

\bibitem[Alameddine et~al., 2023c]{CORSIKA:2023jyz}
Alameddine, J.-M. et~al. (2023c).
\newblock {The particle-shower simulation code CORSIKA 8}.
\newblock {\em PoS}, ICRC2023:310.

\bibitem[Alameddine et~al., 2025]{Alameddine:2024cyd}
Alameddine, J.-M. et~al. (2025).
\newblock {Simulating radio emission from particle cascades with CORSIKA 8}.
\newblock {\em Astropart. Phys.}, 166:103072.

\bibitem[Alameddine et~al., 2020]{Alameddine_2020}
Alameddine, J.-M., Soedingrekso, J., Sandrock, A., Sackel, M., and Rhode, W.
  (2020).
\newblock Proposal: A library to propagate leptons and high energy photons.
\newblock {\em Journal of Physics: Conference Series}, 1690(1):012021.

\bibitem[Albrecht et~al., 2023]{CORSIKA:2023sfd}
Albrecht, J. et~al. (2023).
\newblock {Comparison and efficiency of GPU accelerated optical light
  propagation in CORSIKA 8}.
\newblock {\em PoS}, ICRC2023:417.

\bibitem[Alves et~al., 2023]{CORSIKA:2023nqx}
Alves, A.~A. et~al. (2023).
\newblock {Parallel processing of radio signals and detector arrays in CORSIKA
  8}.
\newblock {\em PoS}, ICRC2023:469.

\bibitem[Ballarini et~al., 2024]{Ballarini:2024uxz}
Ballarini, F. et~al. (2024).
\newblock {The FLUKA code: Overview and new developments}.
\newblock {\em EPJ Nuclear Sci. Technol.}, 10.

\bibitem[Bierlich et~al., 2022]{Bierlich:2022pfr}
Bierlich, C. et~al. (2022).
\newblock {A comprehensive guide to the physics and usage of PYTHIA 8.3}.
\newblock {\em SciPost Phys. Codeb.}, 2022:8.

\bibitem[Coleman et~al., 2024]{Coleman:2024cln}
Coleman, A., Glaser, C., Rice-Smith, R., Barwick, S., and Besson, D. (2024).
\newblock {In-ice Askaryan Emission from Air Showers: Implications for Radio
  Neutrino Detectors}.

\bibitem[De~Kockere et~al., 2024]{DeKockere:2024qmc}
De~Kockere, S., Van~den Broeck, D., Latif, U.~A., de~Vries, K.~D., van
  Eijndhoven, N., Huege, T., and Buitink, S. (2024).
\newblock {Simulation of radio signals from cosmic-ray cascades in air and ice
  as observed by in-ice Askaryan radio detectors}.
\newblock {\em Phys. Rev. D}, 110(2):023010.

\bibitem[Gaudu, 2024]{Gaudu:2024mkp}
Gaudu, C. (2024).
\newblock {Pythia 8 and Air Shower Simulations: A Tuning Perspective}.
\newblock In {\em {22nd International Symposium on Very High Energy Cosmic Ray
  Interactions}}.

\bibitem[Gaudu et~al., 2024]{Gaudu:2024rsq}
Gaudu, C., Reininghaus, M., and Riehn, F. (2024).
\newblock {CORSIKA 8 with Pythia 8: Simulating Vertical Proton Showers}.
\newblock In {\em {28th European Cosmic Ray Symposium}}.

\bibitem[Heck et~al., 1998]{Heck:1998vt}
Heck, D., Knapp, J., Capdevielle, J.~N., Schatz, G., and Thouw, T. (1998).
\newblock {CORSIKA: A Monte Carlo code to simulate extensive air showers}.

\bibitem[Ostapchenko, 2011]{Ostapchenko:2010vb}
Ostapchenko, S. (2011).
\newblock {Monte Carlo treatment of hadronic interactions in enhanced Pomeron
  scheme: I. QGSJET-II model}.
\newblock {\em Phys. Rev. D}, 83:014018.

\bibitem[Ostapchenko, 2024]{Ostapchenko:2024onp}
Ostapchenko, S. (2024).
\newblock {QGSJET-III: predictions for extensive air shower characteristics and
  the corresponding uncertainties}.

\bibitem[Pierog et~al., 2015]{Pierog:2013ria}
Pierog, T., Karpenko, I., Katzy, J.~M., Yatsenko, E., and Werner, K. (2015).
\newblock {EPOS LHC: Test of collective hadronization with data measured at the
  CERN Large Hadron Collider}.
\newblock {\em Phys. Rev. C}, 92(3):034906.

\bibitem[Pierog and Werner, 2023]{Pierog:2023ahq}
Pierog, T. and Werner, K. (2023).
\newblock {EPOS LHC-R : up-to-date hadronic model for EAS simulations}.
\newblock {\em PoS}, ICRC2023:230.

\bibitem[Reininghaus et~al., 2021]{Reininghaus:2021zge}
Reininghaus, M., Ulrich, R., and Pierog, T. (2021).
\newblock {Air shower genealogy for muon production}.
\newblock {\em PoS}, ICRC2021:463.

\bibitem[Riehn et~al., 2020]{Riehn:2019jet}
Riehn, F., Engel, R., Fedynitch, A., Gaisser, T.~K., and Stanev, T. (2020).
\newblock {Hadronic interaction model Sibyll 2.3d and extensive air showers}.
\newblock {\em Phys. Rev. D}, 102(6):063002.

\bibitem[Sandrock et~al., 2023]{Sandrock:2023nfn}
Sandrock, A., Alameddine, J.-M., and Riehn, F. (2023).
\newblock {Validation of Electromagnetic Showers in CORSIKA 8}.
\newblock {\em PoS}, ICRC2023:393.

\bibitem[Zas et~al., 1992]{Zas:1991jv}
Zas, E., Halzen, F., and Stanev, T. (1992).
\newblock {Electromagnetic pulses from high-energy showers: Implications for
  neutrino detection}.
\newblock {\em Phys. Rev. D}, 45:362--376.

\end{thebibliography}

\end{document}